\documentstyle[psfig]{mn}
\title{Peculiar velocities of galaxy clusters: a comparison with
the linear theory}

\author[Mirt Gramann and Ivan Suhhonenko]
{Mirt Gramann and Ivan Suhhonenko
   \\
   Tartu Observatory,
       T\~oravere 61602, Estonia}
\begin{document}
\maketitle

\let\sec=\section
\let\ssec=\subsection
\let\sssec=\subsubsection

\def\kms{\;{\rm km\,s^{-1}}}
\def\kmsmpc{\;{\rm km\,s^{-1}\,Mpc^{-1}}}
\def\hompc{\,h\,{\rm Mpc}^{-1}}
\def\mpcoh{\,h^{-1}\,{\rm Mpc}}
\def\mpc3h{\,h^{3}\,{\rm Mpc^{-3}}}

\begin{abstract}
We investigate peculiar velocities predicted for
clusters in Lambda cold dark matter ($\Lambda$CDM) models
assuming that the initial density fluctuation field is Gaussian.
To study the non-linear regime, we use N-body
simulations. We investigate the rms velocity and the probability
distribution function of cluster peculiar velocities for
different cluster masses. To identify clusters in the simulation
we use two methods: the standard friends-of-friends (FOF) method
and the method, where the clusters are defined as maxima of
a smoothed density field (DMAX). The density field is smoothed with a
top-hat window, using the smoothing radii $R_s=1.5h^{-1}$ Mpc and
$R_s=1.0h^{-1}$ Mpc. The peculiar velocity of the DMAX clusters is defined
to be the mean peculiar velocity of matter within a sphere of the
radius $R_s$. We find that the rms velocity of the FOF clusters
decreases as the cluster mass increases. The rms velocity of the DMAX
clusters is almost independent of the cluster mass and is well
approximated by the linear rms peculiar velocity smoothed at the
radius $R=R_s$. The velocity distribution function of the DMAX clusters
is similar to a Gaussian.

\end{abstract}


\sec{INTRODUCTION}

One of the interesting problems in cosmology is the evolution of
the large-scale peculiar velocity field in the Universe. The
evolution of the large-scale velocity field has been studied in a
number of papers (see, e.g., Davis et al. 1985; Kofman et al.
1994; Brainerd \& Villumsen 1994; Gramann et al. 1995; Jenkins et
al. 1998; Juszkiewicz, Springel \& Durrer 1999; Colin,
Klypin \& Kravtsov 2000; Gramann \& Suhhonenko 2002; Ciecielag et al. 2003).
The linear rms peculiar velocity on a given scale $R$ can be expressed as
$$
\sigma_v(R)=H_0 f \,\, \sigma_{-1} (R), \eqno(1)
$$
where $H_0$ is the Hubble constant and $f$ is the
dimensionless growth rate, which is related to the cosmological matter
density parameter, $\Omega_m$, and the cosmological constant,
$\Omega_{\Lambda}$, by
$$
f(\Omega_m,\Omega_{\Lambda}) \simeq \Omega_m^{0.6} +
{\Omega_{\Lambda} \over 70} \left(1 + {\Omega_m \over 2}\right)
\eqno(2)
$$
(Lahav et al. 1991). The spectral moments $\sigma_j(R)$ are defined for
any integer $j$ by
$$
\sigma_j^2(R)={1 \over 2\pi^2} \int P(k) W^2(kR) k^{2j+2} dk,
\eqno(3)
$$
where $P(k)$ is the power spectrum of density fluctuations and
$W(kR)$ is a Fourier transform of the smoothing window. For the
top-hat window in real space that we will use in this paper,
$W(x)=(3/x^3)[\sin(x)-x\cos(x)]$. Notice that the predicted rms
peculiar velocity depends both on cosmology and on the shape of the
power spectrum.

Jenkins et al. (1998) used a N-body simulation to investigate
the rms peculiar velocity of the dark matter smoothed on the scales
$R=10-80h^{-1}$ Mpc. They compared the results with the linear
approximation (1) and found that on scales above $20h^{-1}$ Mpc, the
linear theory prediction agrees very well with the simulation.
Kofman et al. (1994) and Ciecielag et al. (2003) studied the
evolution of the peculiar velocity one-point distribution
function, beginning with Gaussian initial fluctuations. They
showed that on mildly non-linear scales ($4-10h^{-1}$Mpc,
Gaussian smoothing) the distribution of the Cartesian components
of the peculiar velocity field is well approximated by a
Gaussian. On smaller scales the velocity distribution of dark
matter becomes non-Gaussian due to the motions of matter within
dense systems (see e.g. Sheth \& Diaferio (2001) for the
discussion of the velocity distribution of dark matter particles
in different cosmological models).

Important information for the large-scale velocity field is provided by
peculiar velocities of galaxy clusters. The evolution of peculiar velocities
of clusters in different cosmological models has been examined in
several papers (e.g. Bahcall, Gramann \& Cen 1994; Croft \& Efstathiou 1994;
Moscardini et al. 1996; Suhhonenko \& Gramann 1999; Colberg et al. 2000;
Sheth \& Diaferio 2001; Suhhonenko \& Gramann 2003;
Hamana et al. 2003). Clusters represent high-density maxima in the
dark matter density field. Bardeen et al. (1986) studied the peculiar
velocity distribution for peaks of a Gaussian density field. They showed
that the velocity distribution function for peaks is also
Gaussian, with the rms peculiar velocity
$$
\sigma_p(R)=\sigma_v(R) \sqrt{1 - \sigma_0^4/\sigma_1^2
\sigma_{-1}^2} . \eqno(4)
$$
Peaks have lower peculiar velocities than field points. The
reason for this difference is the fact that the velocity field is
correlated with the density gradient field. However, at radii
$R\sim 1-5h^{-1}$ Mpc, the difference between $\sigma_v(R)$ and
$\sigma_p(R)$ is small in standard cosmological models. For
example, in the flat $\Omega_m=0.3$ $\Lambda$CDM model, it is
about 3 per cent at the radius $R=1.5h^{-1}$ Mpc. Gramann et al. (1995)
studied the large-scale velocity and potential maps of clusters
and of the matter using N-body simulations. They found that
the large-scale velocity and potential fields are recovered remarkably
well by using the velocities of clusters,

In this paper we study peculiar velocities of clusters in a
$\Lambda$CDM model. We use the N-body simulation carried out by
the Virgo Consortium for the flat $\Lambda$CDM model with the
density parameter $\Omega_m=0.3$ (Jenkins et al. 1998). To
identify clusters in the simulation we use two methods: (1) the
standard friends-of-friends (FOF) method and (2) the method,
where the clusters are defined as the maxima of the smoothed
density field (DMAX). We use a top-hat window with the smoothing
radii $R_s=1.5h^{-1}$Mpc and $R_s=1h^{-1}$ Mpc. The velocity and the mass
of FOF clusters are defined to be as mean velocity and the total mass
of all the particles in the cluster. To determine the mass and velocity of
DMAX clusters, we use the same smoothing as for the density
field. We study the rms velocity of clusters for different
cluster masses. The rms peculiar velocity of clusters was studied also
by Suhhonenko \& Gramann (2003). They used a somewhat different DMAX method.
The densities and peculiar velocities of clusters were determined using the
cloud-in-cell (CIC) scheme. In this paper we use a top-hat window, which
allows us to simply calculate cluster masses.

We compare the rms velocity of clusters in the simulation with the
linear approximation (4). What window function, $W(kR)$, and smoothing
length, $R$, for the spectral moments in the equation (4) are appropriate
for cluster velocities? Different studies have used different
window functions and smoothing lengths. Croft \& Efstathiou (1994) and
Bahcall, Gramann \& Cen (1994) used Gaussian smoothing with a radius
$R=3h^{-1}$ Mpc. Suhhonenko \& Gramann (1999) adopted top-hat smoothing
with a radius $R=1.5h^{-1}$Mpc. To compare model predictions with the
observed peculiar velocities of galaxy clusters, Borgani et al. (2000)
used Gaussian smoothing with a radius $R=1.5h^{-1}$ Mpc.
Colberg et al. (2000) and Sheth \& Diaferio (2001) connected the smoothing
length $R$ in the equation (4) with the cluster mass $M$. They studied
the rms peculiar velocity, $\sigma_p(R)$, at the linear radius $R=R_L(M)$,
where the linear radius $R_L$ is defined as $M=4\pi/3 \rho_b R_L^3$
(here $\rho_b$ is the mean background density). Similar approach was used
recently by Hamana et al. (2003).

The velocity of DMAX clusters is determined as the mean velocity of matter
within a sphere of radius $R_s=1.5h^{-1}$ Mpc (or $R_s=1h^{-1}$Mpc). In other
words, we use top-hat smoothing with a radius $R_s$. For this reason, we
adopt also the top-hat window function and the smoothing radius $R=R_s$ in the
equation (4). We consider also the rms peculiar velocity,
$\sigma_p$, at the radius $R=R_L(M)$.

The sizes of FOF clusters are not fixed. If we use the FOF method, the
mean size of high-mass clusters is larger than the mean size of low-mass
clusters. Therefore, it is not possible to connect the rms
velocities of FOF clusters with a single radius $R=R_s$.
We can use the approximation $R=R_L(M)$. Note that for the DMAX
clusters, the ratio $R_L^3/R_s^3=\rho/\rho_b$, where $\rho$
is the mean cluster density within a sphere of radius $R_s$. For
$\rho/\rho_b \approx 200$, the ratio $R_L/R_s \approx 6$.

We also investigate the velocity distribution function of clusters.
We examine the distribution of one-dimensional peculiar
velocities for different cluster masses. Do the clusters exhibit a
Gaussian distribution of peculiar velocities? In addition to studying
cluster velocities at the present time, we also study how cluster
velocities evolve from some early time to the present.

This paper is organized as follows. In Section 2 we describe the
algorithms that have been used to identify clusters and to
determine their masses and velocities. In Section 3 we study the
rms peculiar velocity of clusters for different cluster masses
and compare the results with $\sigma_p(R)$ at different radii.
The velocity distribution function is analyzed in Section 4. In
Section 5 we briefly discuss the evolution of cluster velocities.
A summary and discussion are presented in Section 6.

\sec{SIMULATION OF CLUSTERS}

We study cluster velocities in a N-body simulation carried out by
the Virgo consortium for the flat $\Lambda$CDM model with a
cosmological constant. These simulations are described in
detail by Jenkins et al. (1998). The simulations were created
using an adaptive particle-particle/particle-mesh (AP$^3$M) code
as described by Couchman, Thomas \& Pearce (1995) and Pearce \&
Couchman (1997). In the $\Omega_m=0.3$ $\Lambda$CDM model studied
here, the power spectrum of the initial conditions was chosen to
be in the form given by Bond \& Efstathiou (1984),
$$
P(k)={Ak \over [1+(aq+(bq)^{3/2}+(cq)^2)^\nu]^{2/\nu}}, \eqno(5)
$$
where $q=k/\Gamma$, $a=6.4h^{-1}$Mpc, $b=3h^{-1}$Mpc,
$c=1.7h^{-1}$Mpc, $\nu=1.13$ and $\Gamma=\Omega_0 h = 0.21$. The
normalization constant, $A$, was chosen by fixing the value of
$\sigma_8$ (the linearly extrapolated mass fluctuation in spheres
of radius $8h^{-1}$Mpc) to be $0.9$. The initial density fluctuation
field was assumed to be Gaussian.

The evolution of particles was followed in the comoving box of size
$L=239.5h^{-1}$Mpc. The number of particles was $N_p=256^3$. Therefore,
the mean particle separation $\lambda_p=L/N_p^{1/3}=0.9355 h^{-1}$Mpc
and the mass of a particle $m_p=\rho_b \lambda_p^3=6.82 \times 10^{10} h^{-1}
M_{\odot}$. In the flat $\Omega_m=0.3$ model, the dimensionless growth rate
$f=0.513$ (Peebles 1984).

We used two different algorithms to identify clusters in
simulation: the standard friends-of-friends (FOF) algorithm, and
the algorithm, where clusters are defined as maxima of the
smoothed density field (DMAX).

The friends-of-friends group finder algorithm was applied using
the program suite developed by the cosmology group in the
University of Washington. These programs are available at {\it
http://www-hpcc.astro.washington.edu.} To test our FOF output
data, Suhhonenko \& Gramann (2003) investigated the mass function
of clusters. The cluster mass function in the Virgo
simulations has been studied in detail by Jenkins et al. (2001).
Suhhonenko \& Gramann (2003) found that the agreement between our
results and these obtained by Jenkins et al. (2001) is very good.

The FOF cluster finder depends on one parameter $b$, which
defines the linking length as $b\lambda_p$. The conventional
choice for this parameter is $b=0.2$ (see e.g. G\"otz, Huchra \&
Brandenberger 1998; Jenkins et al. 2001). In this paper we also
define clusters by using the value $b=0.2$. We also study
velocities of the clusters defined by the parameters $b=0.15$. In
the limit of very large numbers of particles per object, FOF
approximately selects the matter enclosed by an isodensity
contour at $\rho_b/b^3$.

We studied FOF clusters that contained at least ten particles. The
three-dimensional peculiar velocity of each cluster was defined as
$$
\vec v_{cl} = {1\over N_f} \sum_{i=1}^{N_{f}} \vec v_i, \eqno(6)
$$
where $N_{f}$ is the number of particles in the cluster and $\vec
v_i$ is the peculiar velocity of the particle $i$ in the cluster.

We also selected DMAX clusters using the following method.

(1) We calculated the density contrast on a grid. For each grid
point, the density contrast was determined as
$$
\delta = {N \over \bar N} -1, \eqno(7)
$$
where $N$ is the number of particles in the sphere of radius $R_s$
around the grid point, and
$$
\bar N = {4 \pi \over 3} {N_p \, R_s^3 \over L^3} \eqno(8)
$$
is the mean number of particles in the sphere of radius $R_s$.

(2) We found the density maxima on the grid. The grid point was
considered as a density maximum, if its density contrast was
higher than the density contrast in all 26 neighbouring grid points.
The location of the grid point, where the density contrast had a
maximum value, was identified as the candidate cluster centre.

(3) The final cluster list was obtained by deleting the candidate
clusters with lower density contrast in all pairs separated by
less than the radius $R_s$.

In this way we define the clusters as maxima of the density field
smoothed by a top-hat window with a radius $R_s$. The smoothing length
sets a lower limit on the size of detected structures in the simulations.

We used the smoothing radii $R_s=1.5h^{-1}$ Mpc and $R_s=1.0h^{-1}$Mpc.
The mean number of particles in spheres of $R_s=1.5h^{-1}$ Mpc and
$R_s=1.0h^{-1}$ Mpc is $\bar N=17.26$ and $\bar N=5.12$,
respectively. To select the $R_s=1.5h^{-1}$ Mpc clusters, we used a
$256^3$ grid (the cell size $l=0.936 h^{-1}$ Mpc). For
$R_s=1.0h^{-1}$ Mpc, we used a $350^3$ grid ($l=0.684 h^{-1}$ Mpc).
For comparison, for the $R_s=1.0h^{-1}$ Mpc clusters we used also a
$256^3$ grid.

We studied the rms density contrast and the rms peculiar velocity
on the $256^3$ grid. The rms density contrast on the grid was
$5.05$ and $7.29$ for the radii $R_s=1.5h^{-1}$ Mpc and
$R_s=1.0h^{-1}$ Mpc, respectively. The rms peculiar velocity,
$\sigma_v$, was determined for the fraction of grid points, $F$,
where the number of particles $N>1$. If there are no particles in
the neighbourhood of a grid point, the velocity field is
undetermined. For $R_s=1.5h^{-1}$ Mpc, we found that $F=0.96$
and $\sigma_v=473 \kms$. For $R_s=1.0 h^{-1}$ Mpc,
$F=0.69$ and $\sigma_v=481 \kms$, respectively. Here we took into
account the finite size of the simulation box (see next section).
We also studied the rms density contrast and the rms peculiar
velocity on the $350^3$ grid and found similar results.

For each DMAX cluster, we investigated the cluster mass, $M$, and
the peculiar velocity, $v_{cl}$, at the radius $R_s$. The mass in
the cluster was determined as $M=N_d \,m_p$, where $N_d$ is the
number of particles in a sphere of radius $R_s$ around the centre
of the cluster. The peculiar velocity of each cluster was defined as
$$
\vec v_{cl} = {1\over N_d} \sum_{i=1}^{N_{d}} \vec v_i, \eqno(9)
$$
where $\vec v_i$ is the peculiar velocity of the particle $i$ in the DMAX
cluster.

Suhhonenko \& Gramann (2003) compared the cluster peculiar velocities
defined by different methods (i.e. the FOF method versus the DMAX method)
for massive clusters. By using different methods to identify the
clusters, we select almost the same objects in the simulation.
But we assign different velocities to the same clusters.

\sec{THE RMS PECULIAR VELOCITY OF CLUSTERS}

To determine the rms peculiar velocities of clusters, we used the
equation
$$
v_{rms}^2= v_s^2 + v_L^2 = {1\over N_{cl}}{\sum_{i=1}^{N_{cl}}
v_{cli}^2} +v_L^2, \eqno (10)
$$
where the parameter $v_s$ describes the dispersion of cluster
velocities, $v_{cli}$, derived from the simulations and the parameter
$v_L$ is the linear contribution from the velocity fluctuations on scales
greater than the size of the simulation box $L$. It is given by
$$
v_L^2 ={H_0^2 \, f^2 \over{2\pi^2}}\int_{0}^{2\pi \over L}P(k)dk.
\eqno (11)
$$
$N_{cl}$ is the number of clusters studied. Using eq. (4),
the linear rms peculiar velocity of peaks can be written as
$$
\sigma_p^2(R)=\sigma_v^2(R) - H_0^2 \, f^2 {\sigma_0^4(R) \over
\sigma_1^2(R)}  . \eqno(12)
$$
The second term in this expression is not sensitive to the
amplitude of large-scale fluctuations at wavenumbers $k<2\pi/L$.
Therefore, the linear rms velocity of peaks can be expressed,
approximately, as
$$
\sigma_p^2(R) \approx \sigma_p^{\prime 2}(R) +v_L^2, \eqno(13)
$$
where $\sigma_p^{\prime} (R)$ is determined by the power spectrum
at the wavenumbers $k>2\pi/L$ and $v_L$ is given by eq. (11). For
the $\Lambda$CDM model studied here, we found that $v_L = 220 \kms$.

As a first step, the one-dimensional distribution of cluster
velocities can be approximated as a Gaussian distribution (see
next section for the study of the velocity distribution of
clusters). If the one-dimensional velocities of clusters, $v_{xi}$,
follow a Gaussian distribution with a mean $\bar v_{x}=0$ and a
dispersion $\sigma^2$, then the sum
$$
\chi^2={1 \over \sigma^2} {\sum_{i=1}^{N_{cl}} v_{cli}^2}
\eqno(14)
$$
is distributed as a $\chi^2$ distribution with the number of
degrees of freedom $\nu=3 N_{cl}$. In this case, the rms error
for the variable $v_{rms}^2$ can be determined as
$$
\Delta v_{rms}^2 = \sqrt {2 \over 3 N_{cl}} \,\, v_s^2. \eqno(15)
$$
We used eq. (15) to estimate the error bars for the rms
velocities of clusters.

The clusters were divided into subgroups according to their mass.
We studied the rms peculiar velocity of clusters in subgroups for
which the mass
was in the ranges $(10^{12} - 5 \times 10^{12})h^{-1}M_{\odot}$, ...,
$(5 \times 10^{14} - 10^{15}) h^{-1} M_{\odot}$. For the DMAX clusters
defined with $R_s=1.5h^{-1}$ Mpc, we considered the clusters with
masses $M > 5 \times 10^{12} h^{-1} M_{\odot}$ ($N/\bar N >4.24$).
Table~1 shows the number of clusters and Fig.~1 demonstrates
the rms peculiar velocity of clusters in different mass intervals.
The rms velocities are shown for the intervals for which the numbers of
clusters are $N_{cl}>10$.

\begin{table}
\caption{The number of clusters, $N_{cl}$, in different mass
intervals.}

\begin{tabular}{|c|r|r|r|r|}
\hline \hline
   $M$  & DMAX & DMAX & FOF & FOF \\
  ($h^{-1}M_{\odot}$) & $R_s=1.5$ & $R_s=1.0$ & b=0.2 & b=0.15 \\
\hline
 $10^{12}$ -- $5\times 10^{12}$ &  ---  & 70153 & 51841 & 47056  \\
 $5\times 10^{12}$ -- $10^{13}$ & 10647 &  9989 &  7326 &  6764  \\
 $10^{13}$ -- $5\times 10^{13}$ & 10049 &  8866 &  6542 &  5720  \\
 $5\times 10^{13}$ -- $10^{14}$ &  1261 &  1052 &   856 &   695  \\
 $10^{14}$ -- $5\times 10^{14}$ &   688 &   470 &   511 &   393  \\
 $5\times 10^{14}$ -- $10^{15}$ &    23 &     4 &    38 &    24  \\
\hline
\end{tabular}
%\label{table}
\end{table}

\begin{figure}
\centering \leavevmode \psfig{file=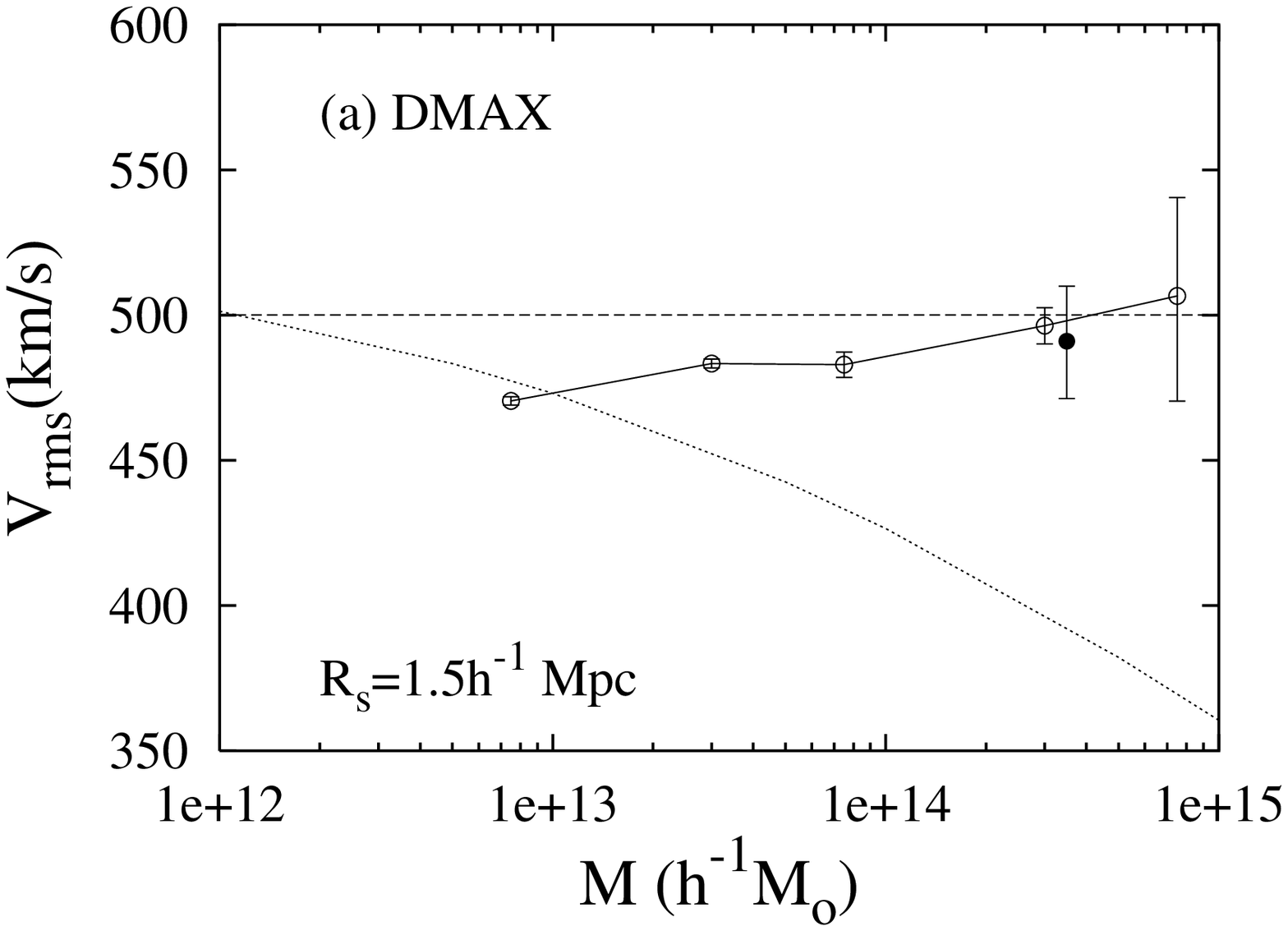,width=8cm}
\psfig{file=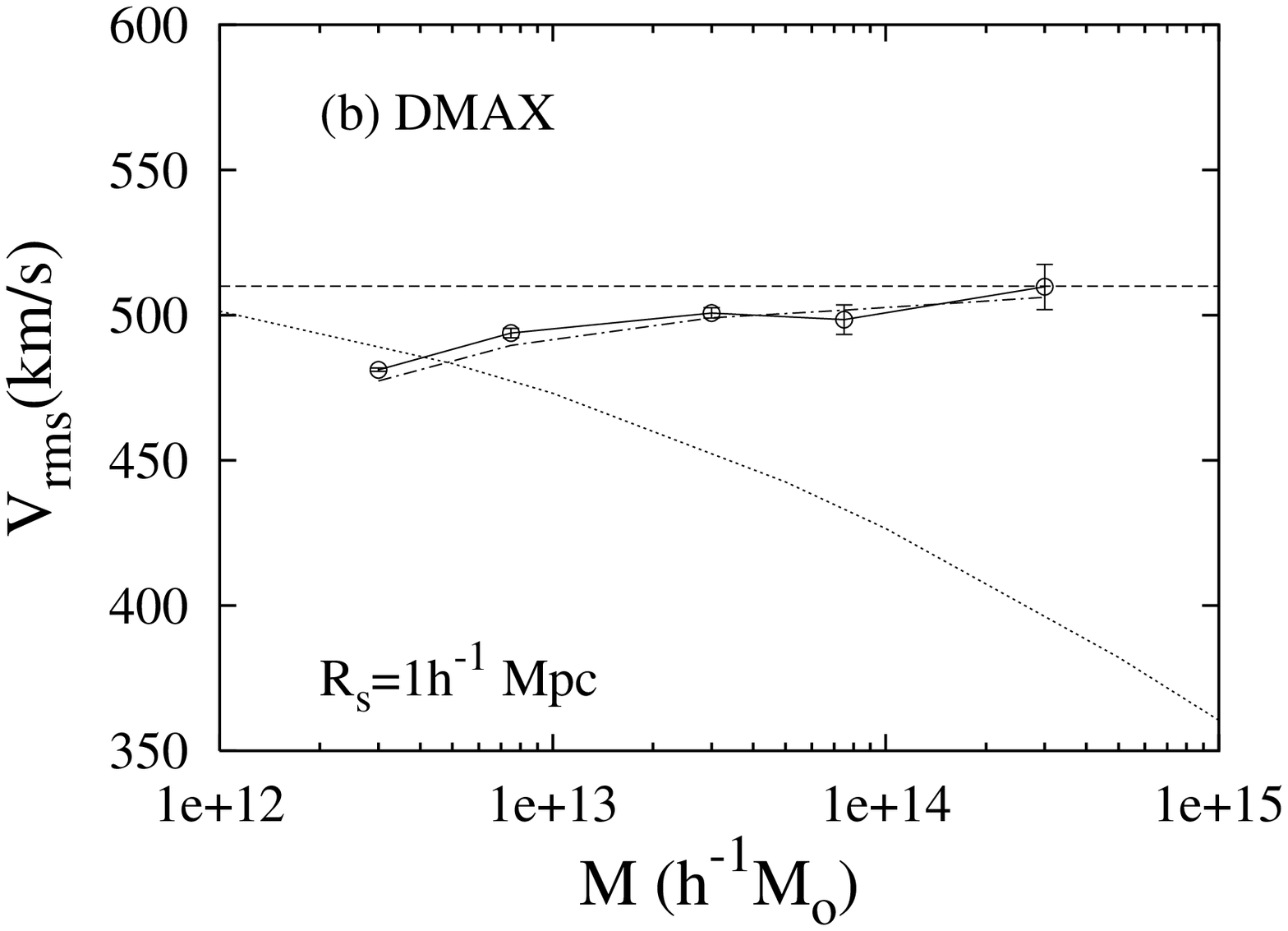,width=8cm} \psfig{file=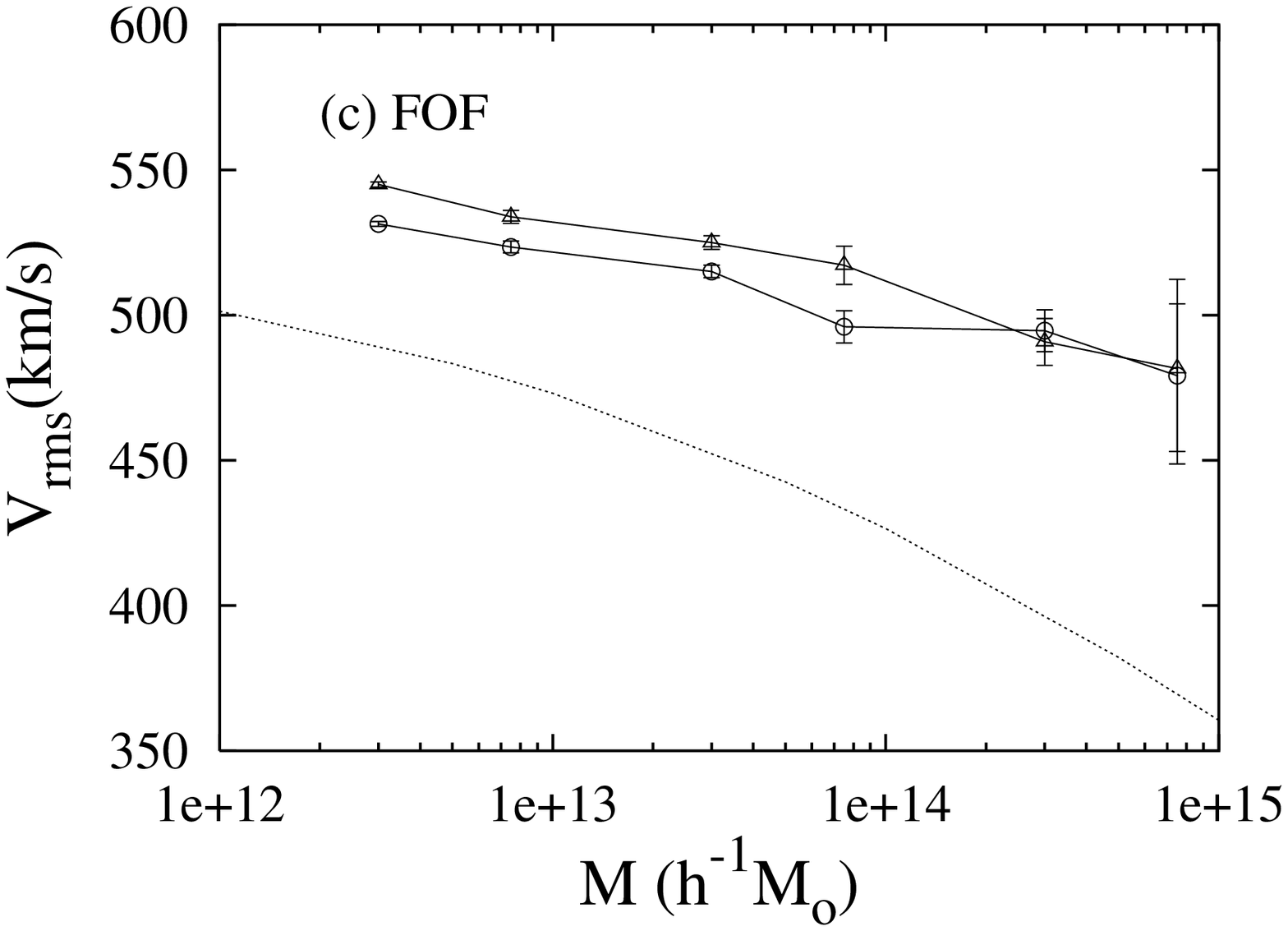,width=8cm}

\caption{The rms peculiar velocities of clusters for different
cluster masses. (a) The velocities for the DMAX clusters defined
with $R_s=1.5h^{-1}$ Mpc (open circles). The solid circle
represents the result obtained by Colberg et al. (2000). The
dashed line shows the linear rms peculiar velocity of peaks,
$\sigma_p$, for the radius $R=1.5h^{-1}$ Mpc. (b) The velocities
for the $R_s=1.0h^{-1}$ Mpc DMAX clusters. The circles show the
velocities for the clusters defined using a $350^3$ grid. The
dot-dashed line describes the cluster velocities for a $256^3$
grid. For comparison, we show the linear rms peculiar velocity of
peaks for the radius $R=1.0h^{-1}$ Mpc (dashed line). (c) The
velocities of the FOF clusters determined by $b=0.2$ (circles)
and $b=0.15$ (triangles). The dotted line in each panel shows the
linear rms peculiar velocity of peaks for the radius $R=R_L(M)$.}

\end{figure}

Fig.1a and Fig~1b demonstrate the rms peculiar velocities for the
DMAX clusters. Fig.~1a shows the results for the radius
$R_s=1.5h^{-1}$ Mpc and Fig.~1b for the radius $R_s=1.0h^{-1}$
Mpc. We see that the rms velocity of DMAX clusters is almost
independent of the cluster mass. The rms velocity of clusters
somewhat increases as the cluster mass increases. However, this
increase is small ($\approx 7$ per cent). For a given scale $R_s$,
the rms velocity of clusters is similar to the rms velocity
for the full grid. The rms velocity of $R_s=1.5h^{-1}$ Mpc
clusters is $471 \kms$ in the mass interval
$(5 \times 10^{12}- 10^{13}) h^{-1} M_{\odot}$ and $507 \kms$ in the
mass interval $(5 \times 10^{14} - 10^{15}) h^{-1} M_{\odot}$.
The rms velocity of $R_s=1.0h^{-1}$ Mpc clusters is $495 \kms$ in
the mass interval $(5 \times 10^{12}- 10^{13}) h^{-1} M_{\odot}$
and $510 \kms$ in the mass interval $(10^{14} - 5 \times 10^{14})
h^{-1} M_{\odot}$. The rms velocities for the $R_s=1.5h^{-1}$ Mpc
clusters are smaller than the rms velocities for the
$R_s=1.0h^{-1}$ Mpc clusters.

The open circles in Fig~1b show the results for the clusters
defined by using the $350^3$ grid. For comparison, we plotted also
the rms velocities for the clusters defined by using the $256^3$
grid. We see that the rms peculiar velocities of clusters in
different mass intervals are not sensitive to the number of cells
in the grid that was used to find the cluster centres.

Our results are in good agreement with the results obtained
Colberg et al. (2000). They also studied the rms velocity of
clusters in the $\Lambda$CDM model, but used a
slightly different method to select clusters. High-density
regions were located using a FOF method with $b=0.05$ and their
barycentres were considered as candidate cluster centers. Any
candidate centre for which mass within $1.5h^{-1}$ Mpc exceeded the
threshold mass $M_t$ was identified as a candidate cluster. The
final cluster list was obtained by deleting the lower mass candidate
in all pairs separated by less than $1.5h^{-1}$ Mpc. The peculiar
velocity of each cluster was defined to be the mean peculiar
velocity of all the particles within the $1.5h^{-1}$ Mpc sphere.
This method to determine the cluster velocities is similar to the
DMAX method, if we use the radius $R_s=1.5h^{-1}$ Mpc.

In Fig.~1a we show the result obtained by Colberg et al. (2000)
for the $\Lambda$CDM model. They used the value $M_t=3.5 \times
10^{14} h^{-1} M_{\odot}$. For this value, the number of clusters
was $N_{cl}=69$. They found that the rms cluster velocity derived
from simulation is $v_s=439 \kms$. If we include the dispersion
$v_L^2$, we find that this value of $v_s$ corresponds to the rms
velocity $v_{rms}=491 \kms$. For comparison, we also studied the
$R_s=1.5h^{-1}$Mpc clusters with masses $M>3.5 \times 10^{14}
h^{-1} M_{\odot}$. We found $72$ clusters with the rms velocity
$v_{rms}=482^{+18}_{-19} \kms$. This value is in good agreement
with the rms velocity found by Colberg et al. (200).

For comparison, we show in Fig.~1a and Fig.~1b the rms peculiar velocity of
peaks, $\sigma_p(R)$, for the radius $R=R_s$.
For the radii $R=1.5h^{-1}$ Mpc and $R=1.0h^{-1}$ Mpc, the $\sigma_p=500 \kms$
and $\sigma_p=510 \kms$, respectively. We see that the rms
velocity of DMAX clusters is well approximated by the linear rms
velocity smoothed at the radius $R=R_s$ (i.e. at the radius which
is used to define the cluster velocity). The rms velocity of
low-mass clusters is somewhat smaller than predicted by the
linear theory.

In Fig.~1c we show the rms peculiar velocities of FOF clusters
determined by $b=0.2$ and $b=0.15$. The rms velocity of FOF
clusters decreases, as the mass of the clusters increases. This
result is in agreement
with the results obtained by Sheth \& Diaferio (2001) and
Hamana et al. (2003). They studied the rms velocities of FOF clusters in
different mass intervals and found that the rms cluster velocity decreases
with mass. Suhhonenko \& Gramann (2003) studied the velocities of FOF
clusters for different masses and radii. They showed that the effect of
the cluster radius on $v_{rms}$ is similar to the effect of the cluster
mass on $v_{rms}$. The rms velocity of small FOF clusters is higher
than the rms velocity of large massive clusters.

We also plot the linear theory rms peculiar velocity of peaks, $\sigma_p$,
defined by the linear radius $R=R_L$ (dotted lines in Fig.~1). In this
approximation, the rms peculiar velocity decreases as the cluster mass
increases. For the mass $M=10^{14} h^{-1} M_{\odot}$, the linear radius
$R_L=6.59h^{-1}$ Mpc and the rms velocity
$\sigma_p(R_L)=427 \kms$. For $M=5 \times 10^{14} h^{-1} M_{\odot}$,
we find that $R_L=11.3 h^{-1}$ Mpc and $\sigma_p(R_L)=382 \kms$. These
values for the rms velocities are significantly lower than the rms velocities
of clusters found from the simulation. This result was first obtained
by Colberg et al. (2000), who compared the cluster velocities with the
linear rms velocities of peaks at the radius $R=R_L$.

\sec{VELOCITY DISTRIBUTION FOR THE CLUSTERS}

Fig.2 and Fig.3 show the velocity distribution function
for the DMAX clusters. We investigated the distribution of one-dimensional
cluster peculiar velocities, ($v_x$,$v_y$,$v_z$), for different cluster
masses. Fig.2 shows the results for the radius $R_s=1.5h^{-1}$ Mpc and
Fig.3 for the radius $R_s=1.0h^{-1}$ Mpc. The probability density
was estimated as the normalized number of clusters in the range
$v_1 \pm \Delta v_1$, as a function of $v_1$. We used the value
$\Delta v_1=100 \kms$. We also show the Poisson error bars for the
velocity distribution.

\begin{figure}
\centering \leavevmode \psfig{file=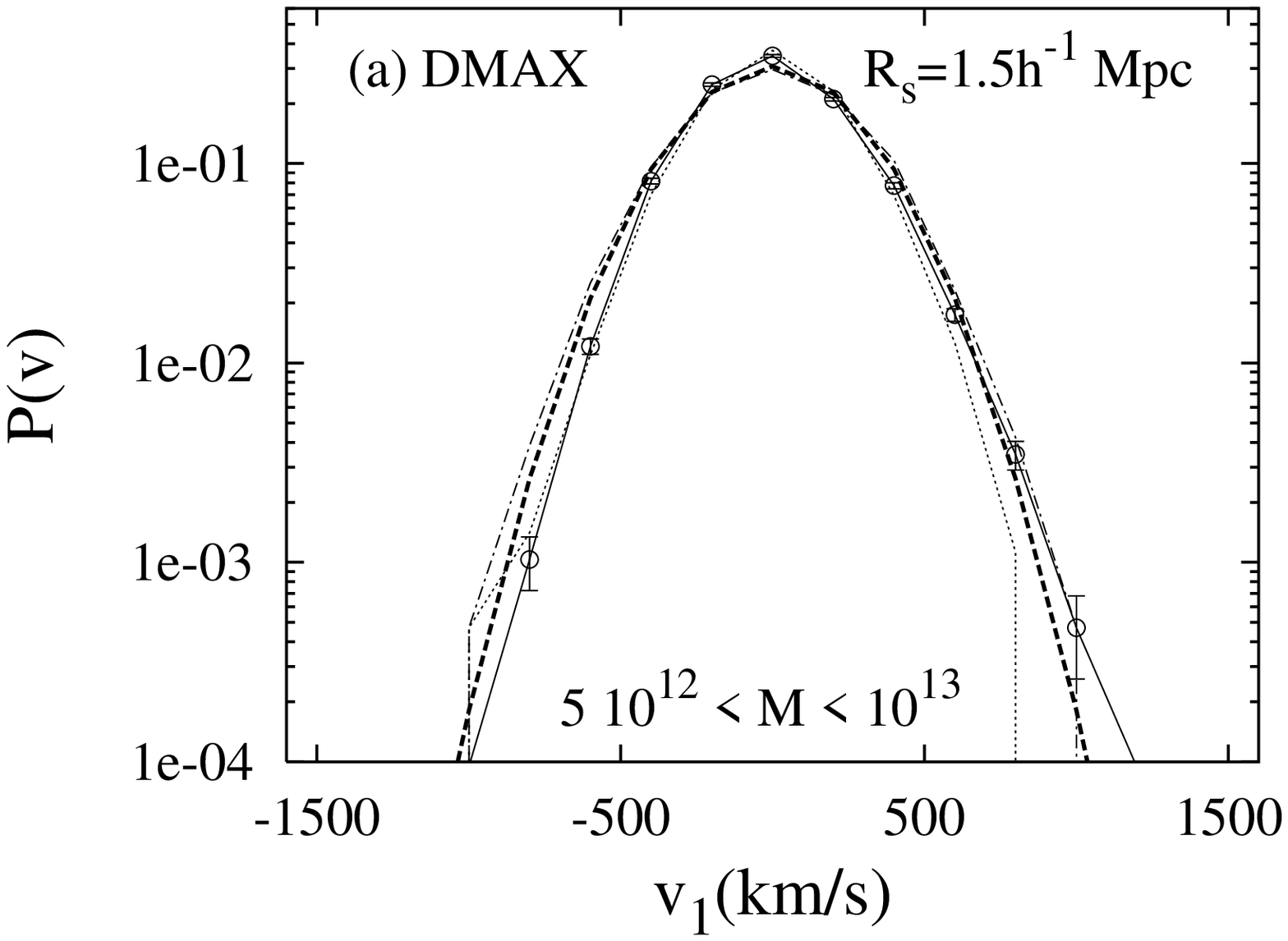,width=8cm}
\psfig{file=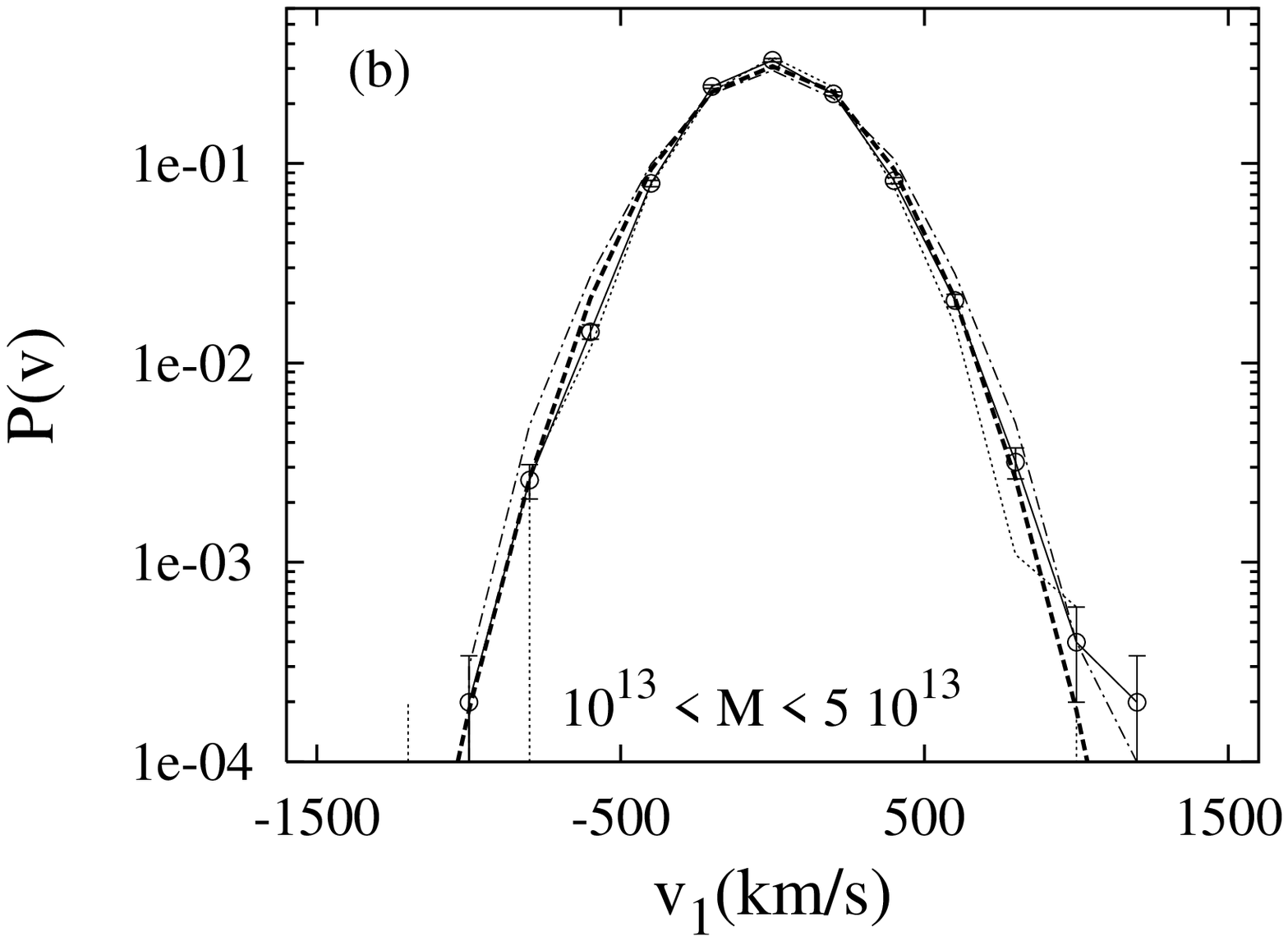,width=8cm} \psfig{file=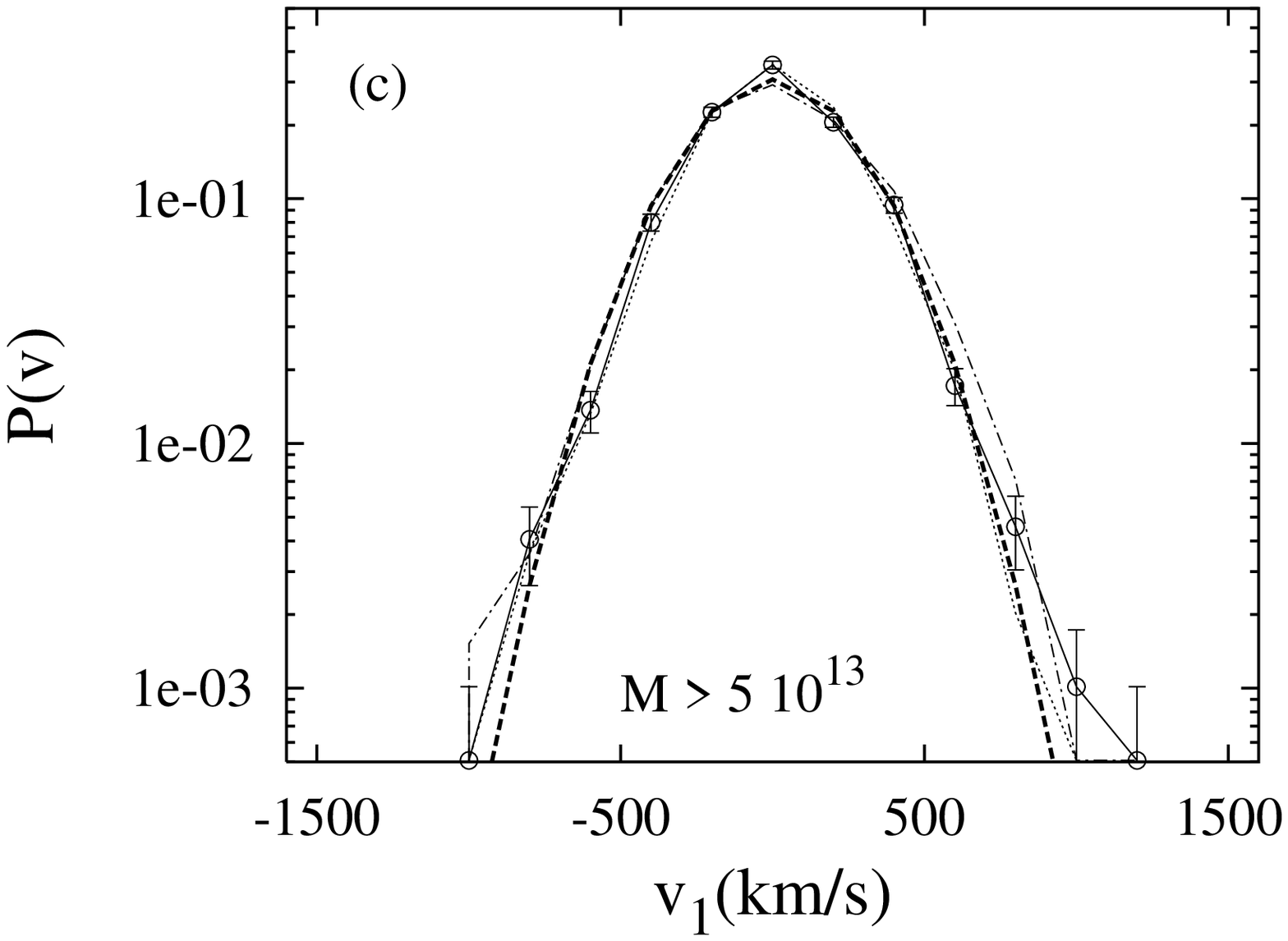,width=8cm}

\caption{Probability densities of one-dimensional
cluster velocity $v_1$ for limited mass ranges (denoted in each
panel). Clusters are defined by the DMAX method with the radius
$R_s=1.5h^{-1}$ Mpc. The solid, dot-dashed and dotted lines show
the distribution of the three Cartesian components
($v_x$,$v_y$,$v_z$), respectively. For clarity, the error bars
are shown only for the $v_x$ distribution. The heavy dashed line
in each panel describes the Gaussian distribution function
predicted by the linear theory for the scale $R=1.5h^{-1}$ Mpc
($\sigma_1=260 \kms$).}

\end{figure}

\begin{figure}
\centering \leavevmode \psfig{file=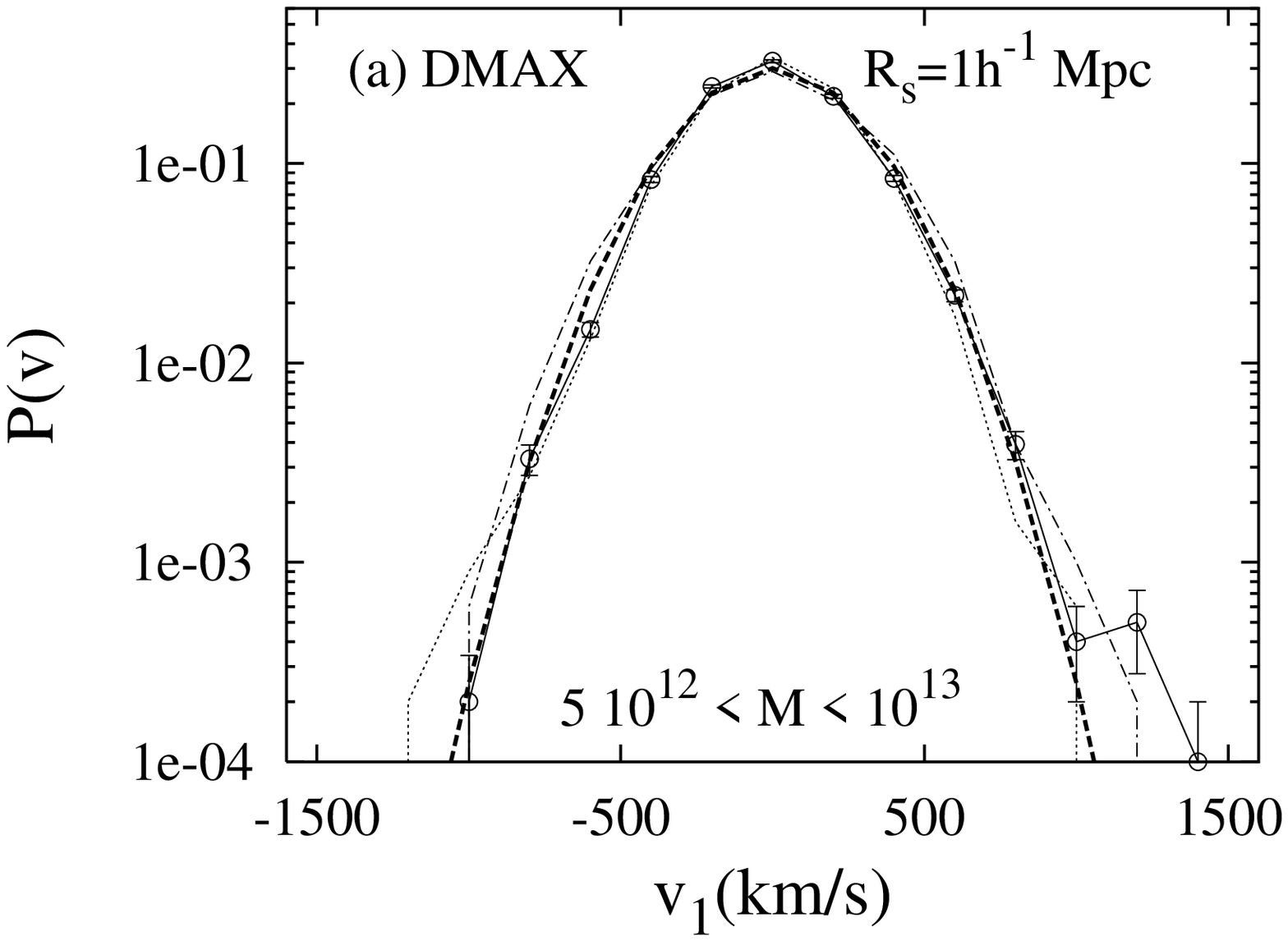,width=8cm}
\psfig{file=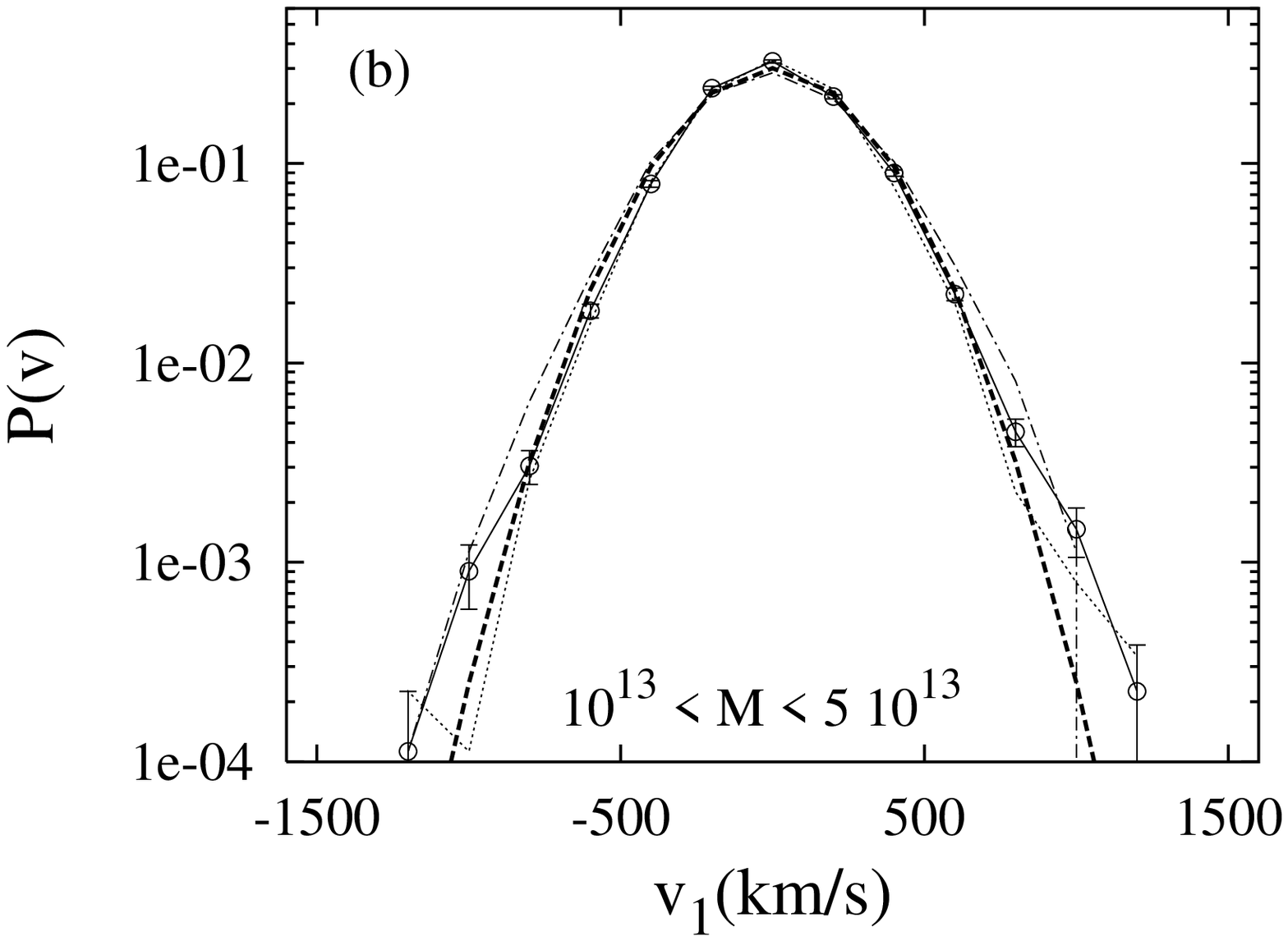,width=8cm} \psfig{file=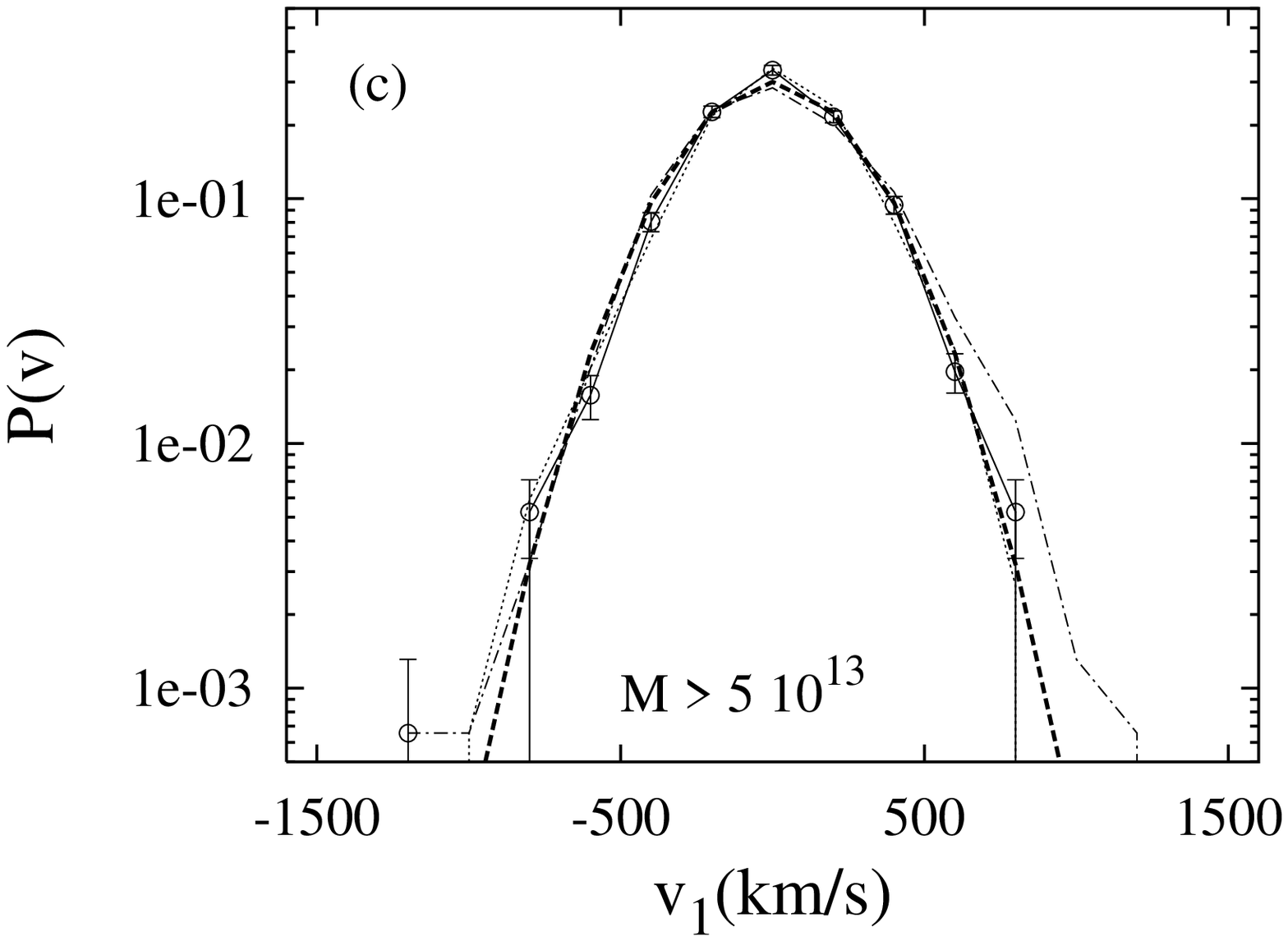,width=8cm}

\caption{As for the previous figure, but for the DMAX clusters
defined with the radius $R_s=1.0h^{-1}$ Mpc. The heavy dashed
line in each panel describes the Gaussian distribution function
predicted by the linear theory for the scale $R=1.0h^{-1}$ Mpc
($\sigma_1=265 \kms$).}

\end{figure}
We examined the velocity distribution for the clusters in three
different mass intervals: (1) $M=(5\times 10^{12} - 10^{13})
h^{-1} M_{\odot}$ (Fig.~2a and Fig.~3a), (2) $M=(10^{13} - 5
\times 10^{14}) h^{-1} M_{\odot}$ (Fig.~2b and Fig.~3b) and (3)
$M>10^{13} h^{-1} M_{\odot}$ (Fig.~2c and Fig.~3c). The number of
clusters, $N_{cl}$, in the mass intervals (1) and (2) is shown in
the Table~1. The number of clusters with masses $M>10^{13} h^{-1}
M_{\odot}$ is $N_{cl}=1973$ and $N_{cl}=1526$ for the radius
$R_s=1.5h^{-1}$ Mpc and $R_s=1.0h^{-1}$ Mpc, respectively. We see
that the velocity distribution function for different mass
intervals is similar.

For comparison, we plot in Fig.2 the Gaussian distribution
with the dispersion $\sigma_1^2$ given by
$\sigma_1=\sigma_p^{\prime}/\sqrt{3}=260 \kms$. This dispersion
is predicted by the linear theory for the scale $R=1.5h^{-1}$ Mpc,
after accounting for the finite size of the simulation box (see eq.
13). At the radius $R=1.5h^{-1}$ Mpc, the rms velocity of peaks
$\sigma_p=500 \kms$ and, therefore, the parameter
$\sigma_p^{\prime}=450 \kms$. Fig.~2 demonstrates that the
one-dimensional velocity distribution function for $R_s=1.5h^{-1}$ Mpc
clusters is well approximated by a Gaussian.

In Fig.3 we compare the velocity distribution of $R_s=1.0h^{-1}$ Mpc
clusters with the Gaussian distribution for $\sigma_1=265 \kms$.
At the radius $R=1.0h^{-1}$ Mpc, the rms velocity of peaks
$\sigma_p=510 \kms$ and, therefore, $\sigma_p^{\prime}=460 \kms$.
We see that the velocity distribution of $R_s=1.0h^{-1}$ Mpc
clusters is also similar to the Gaussian distribution. However,
there are small deviations from the Gaussian distribution,
especially for the clusters with masses $M= (10^{13}- 5\times
10^{13}) h^{-1} M_{\odot}$.

During the evolution larger and larger scales become non-linear
and deviations from the Gaussian distribution develop. As
demonstrated by Bachall, Gramann \& Cen (1994) and
Sheth and Diaferio (2001), evolution of the cluster peculiar
velocities depends on the large-scale density field. The highest
velocity clusters frequently originate in dense superclusters.
Fig.~3 demonstrates how the deviations from the Gaussian distribution
start to develop.

\begin{figure}
\centering \leavevmode \psfig{file=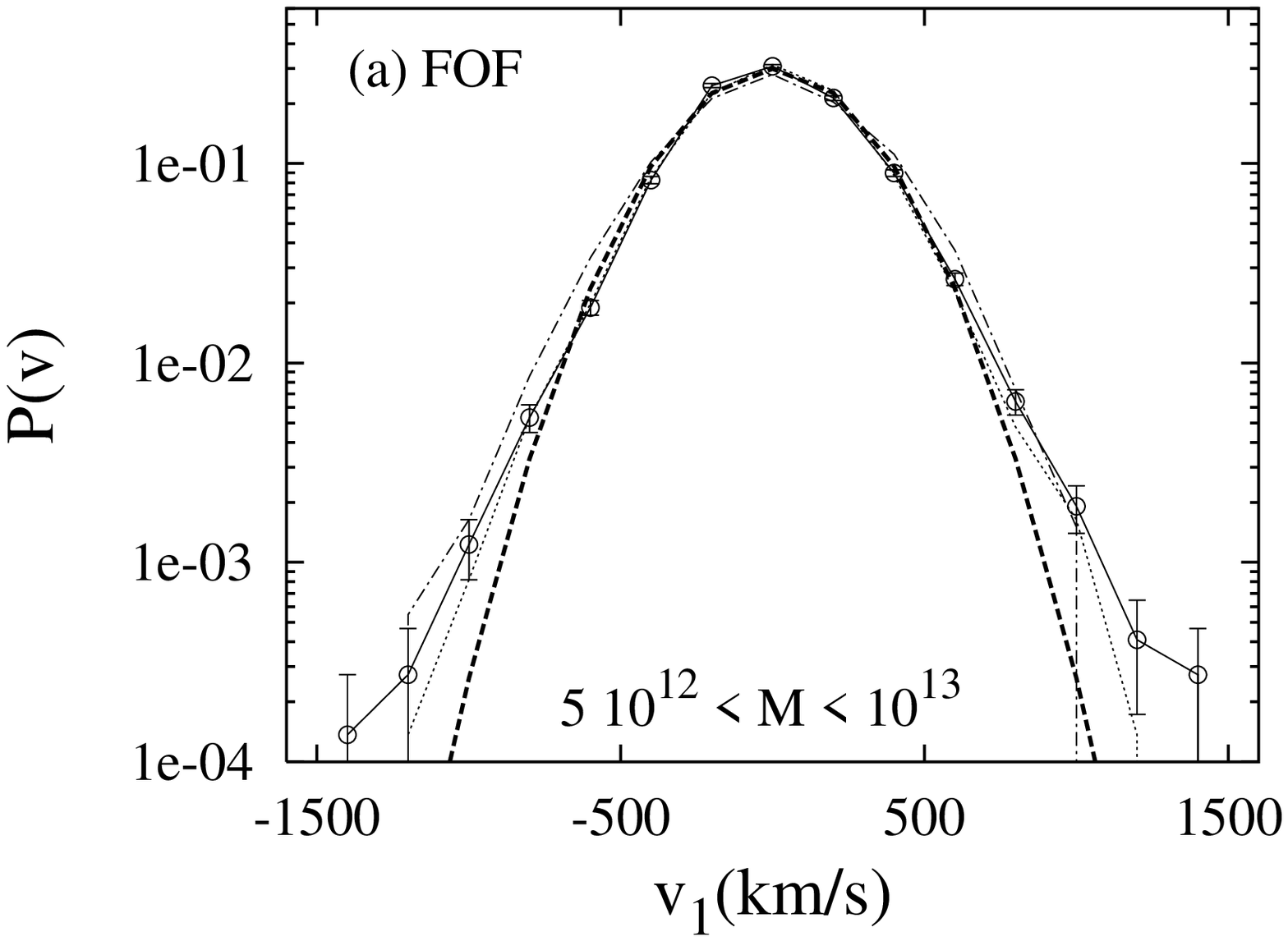,width=8cm}
\psfig{file=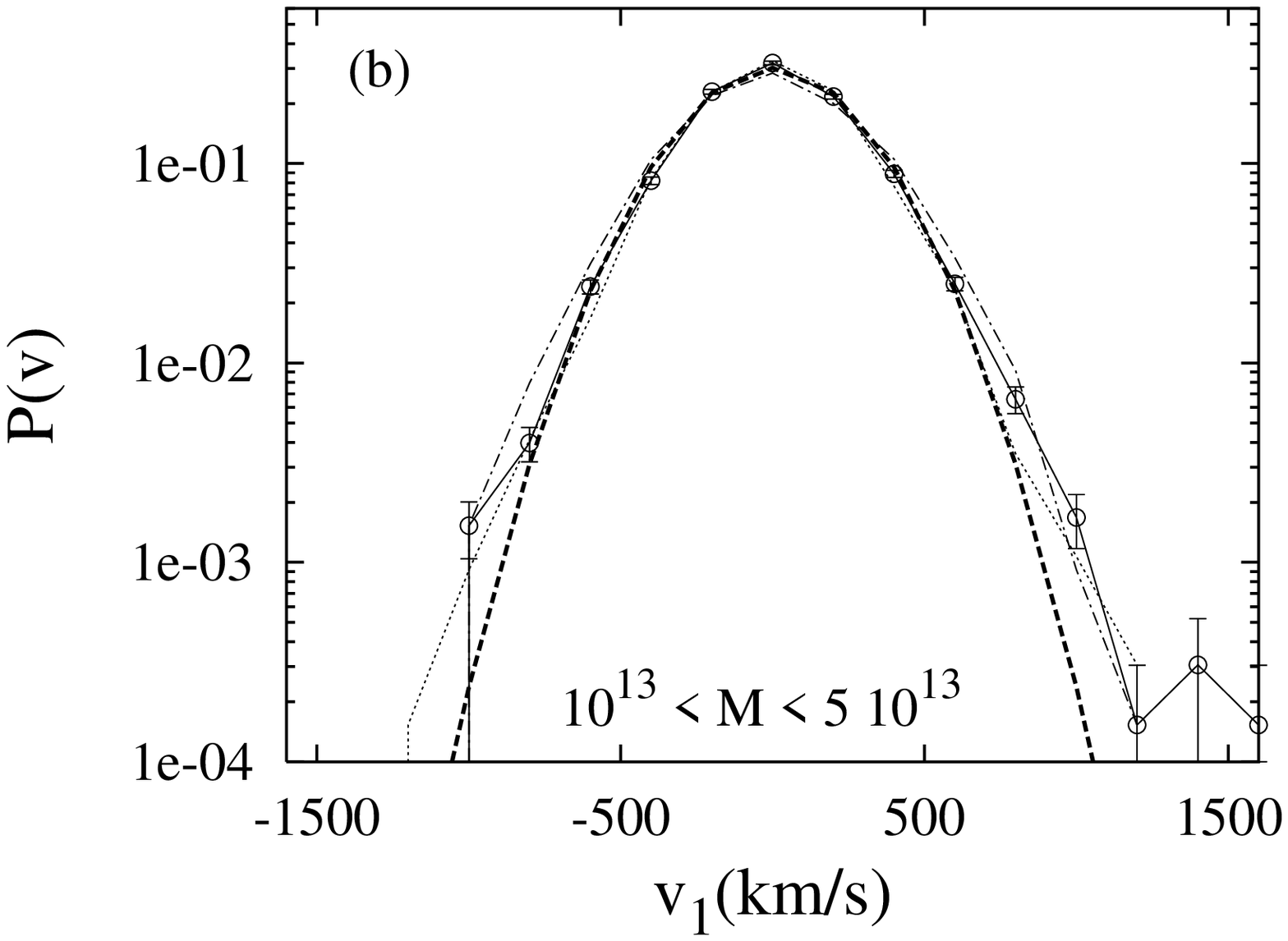,width=8cm} \psfig{file=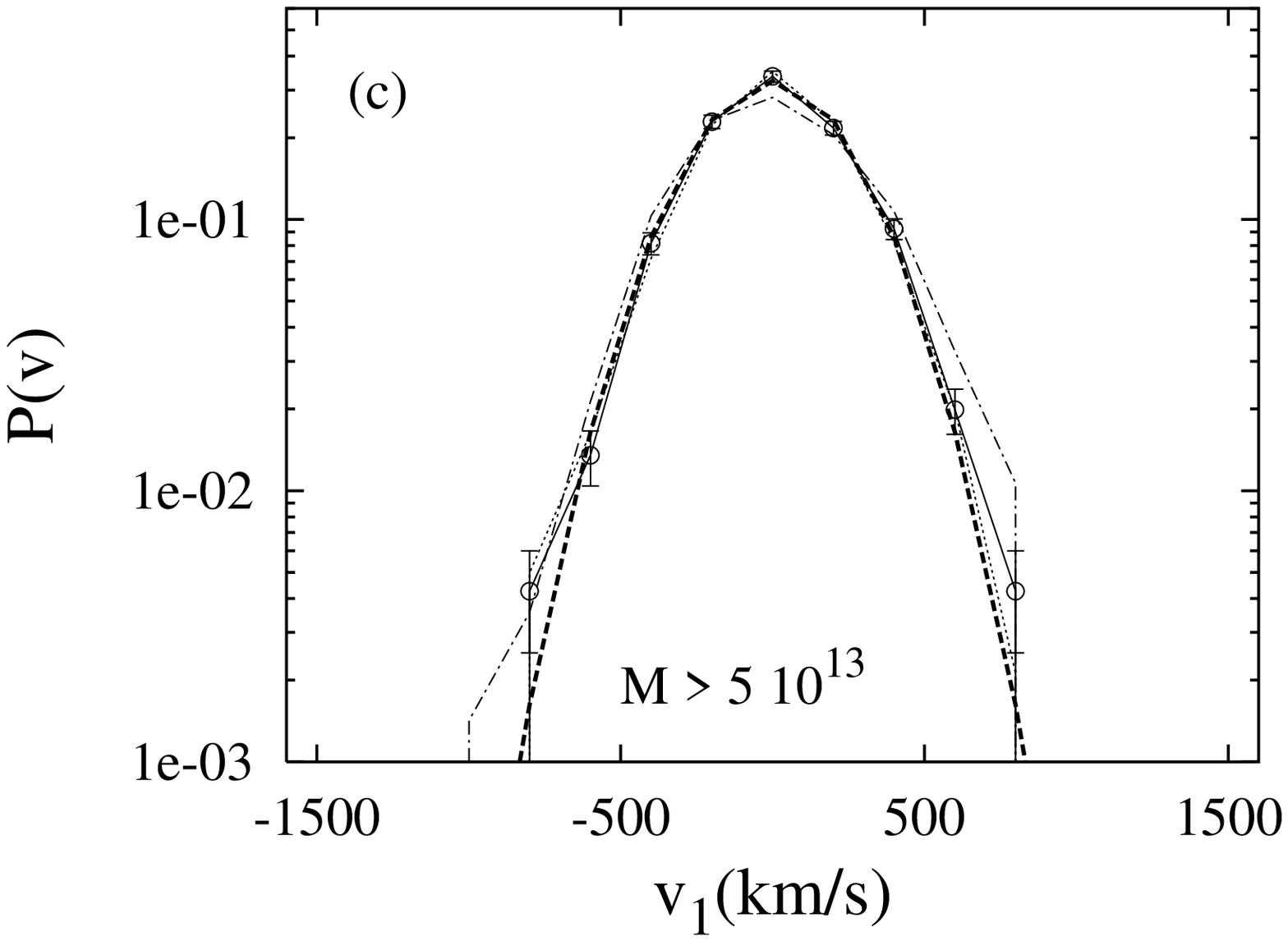,width=8cm}

\caption{As for the figures Fig.2 and Fig.3, but for the FOF clusters
determined by $b=0.2$. The heavy dashed line in each panel describes
the Gaussian distribution function with the dispersion $\sigma_x^2$ as
computed for the $v_x$ velocities in the N-body simulation.}

\end{figure}

Fig.~4 shows the velocity distribution function for the FOF clusters
determined by $b=0.2$. We studied also the velocity distribution
for the clusters defined by $b=0.15$ and found similar results.
The velocity distribution was determined in different mass intervals.
For comparison, we plot the Gaussian distribution function with the
dispersion $\sigma_x^2$ as determined for the $v_x$ velocities in the
simulation. The parameter $\sigma_x=266$, $264$ and $246 \kms$ in the
Fig.~4a, Fig.~4b and Fig.~4c, respectively.

The velocity distribution of FOF clusters in different mass intervals is
different. In Fig~4a and Fig.~4b, we see that the peculiar velocity
distributions found for the clusters have stronger peculiar
velocity tails than predicted by a Gaussian distribution.
Deviations from the Gaussian distribution are larger for small
low-mass clusters. This result for the FOF clusters was obtained
also by Hamana et al. (2003), who studied the velocity
distribution for clusters with masses $M<3.2 \times 10^{13}
h^{-1} M{\odot}$ (see Fig.5 in their study). We also investigated
the velocity distribution for massive clusters with
$M>5 \times 10^{13} h^{-1} M_{\odot}$ (Fig.~4c). The velocity
distribution of massive FOF clusters is similar to the Gaussian
distribution.

\sec{EVOLUTION OF CLUSTER VELOCITIES}

We investigated the evolution of cluster velocities by using the Virgo
simulation for the $\Lambda$CDM model at redshift $z=10$.

We selected $509$ clusters with masses $M>1.2 \times 10^{14}
h^{-1} M_{\odot}$ at $z=0$. The mean intercluster separation in
this sample was $d_{cl}=30h^{-1}$ Mpc. The rms velocity of the
clusters was $v_{rms}=494 \kms$. The clusters were determined with
the DMAX method, using the radius $R_s=1.5h^{-1}$ Mpc. For each
cluster, we found all particles within a $1.5h^{-1}$ Mpc sphere
around the centre of the cluster and determined the positions of
these particles at $z=10$. We define the initial centre of the
cluster at $z=10$ to be the barycentre of these particles.

Are these initial centres of the clusters associated with peaks
in the density field at $z=10$? We studied the density field in
the Virgo simulation at $z=10$. To identify the peaks in the
density field, we used the DMAX method with the radius
$R_s=1.5h^{-1}$ Mpc (comoving). The rms density contrast on the
$256^3$ grid was $\sigma_0=0.31$. For each peak, we determined the
dimensionless height of a peak, $\nu$, and the peculiar velocity
of a peak, $v_p$. The parameter $\nu$ was defined as
$$
\nu = {1 \over \sigma_0} \, \left({N_d \over \bar N} -1 \right),
\eqno(16)
$$
where $N_d$ was the number of particles in the sphere of radius
$R_s$ around the centre of the peak and $\bar N$ was the mean
number of particles in this sphere (see eq. 8). The peculiar
velocity of a peak was defined to be the mean velocity of
particles in the sphere of radius $R_s$. The velocities of the
peaks were scaled up to the values expected at $z=0$ according to
the linear theory. In the linear regime, the peculiar velocity $v
\sim a D H f$, where $a$ is the scale factor and $D$ is the
linear growth factor. By taking into account the evolution of
these four functions, we found that $v(z=0) = 2.42 \, v(z=10)$.
In the following we consider only peaks with $\nu>3$. The rms
peculiar velocity of these peaks was $516 \kms$ (scaled up to
$z=0$).

We investigated the peaks within a $3.0h^{-1}$ Mpc sphere
around the initial centre of the clusters. We found that $413$
clusters ($81$ per cent) are associated with a peak or peaks in
the density field at $z=10$. Most of the remaining clusters can
be associated with a peak of either a slightly lower height or at
a slightly greater separation. For $205$ clusters, we found
one peak at the initial centre of the cluster. $144$ clusters
were associated with two peaks. For $64$ clusters, we found
three or more peaks.

\begin{figure}
\centering \leavevmode \psfig{file=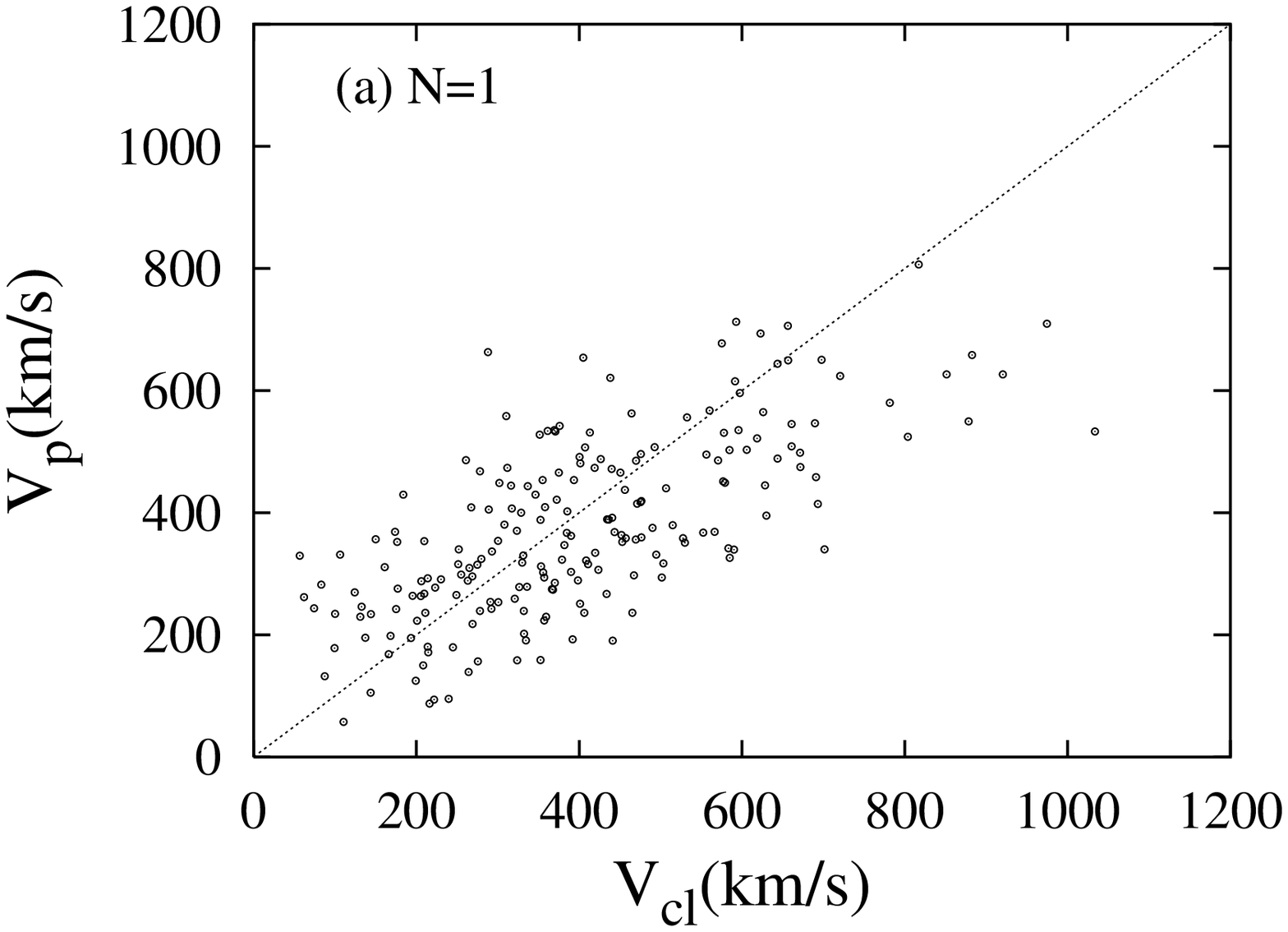,width=8cm}
\psfig{file=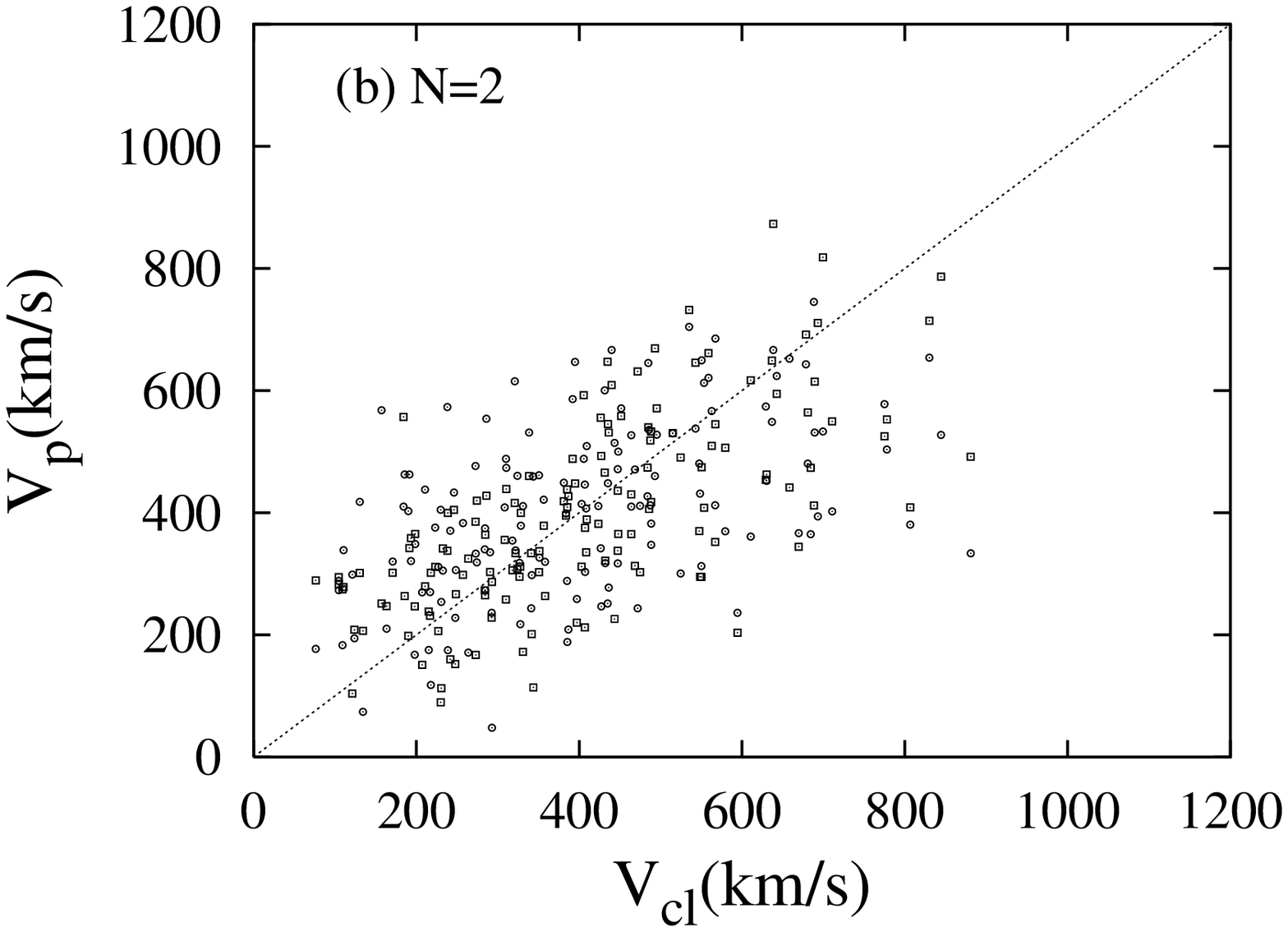,width=8cm}

\caption{Peculiar velocities of peaks at the initial centre of
the clusters, $v_p$, as compared with the final peculiar
velocities of clusters, $v_{cl}$, at $z=0$. The velocities of
peaks are scaled up to values expected at $z=0$ according to
the linear theory. (a) Velocities for the $205$ clusters associated
with one peak. (b) Velocities for the 144 clusters associated
with two peaks. Different symbols in the panel (b) show the
velocities for the different peaks.}

\end{figure}

In Fig.~5 we compare the peculiar velocities of peaks at the
initial centres of clusters with the final peculiar velocities
of clusters at $z=0$. In Fig.~5a we show the velocities for the
clusters associated with one peak. Fig.~5b describes the
results for the clusters associated with two peaks. We see the
correlation between the peculiar velocities of clusters and their
associated peaks. In Fig.~5a, the rms difference in
(3-D) peculiar velocity between a cluster and its associated peak is
$246 \kms$. In Fig. ~5b, the rms difference is $262 \kms$.

At $z=0$, the density field is highly non-linear at the scale
$R=1.5h^{-1}$ Mpc. The rms density contrast is $5.03$. We cannot
consider the density maxima in this field as separate objects,
which move with linear speeds. During the evolution many density
maxima merge and lose their identity. Some clusters obtain
higher velocities than the values expected according to the linear
scaling, other clusters move at lower velocities.
However, these effects compensate each other. Our results show
that the rms peculiar velocity of density maxima at $z=0$ is well
approximated by the linear rms velocity expected on a given scale.
The rms peculiar velocity of density maxima is similar to the rms
peculiar velocity for all field points at the same scale. In other
words, the rms peculiar velocity does not depend much on the
location of spheres used to determine the velocities. In
comparison with the density field, the velocity field is more
heavily weighted by modes with low values of the wavenumber $k$,
i.e. large scales which are in the linear regime.

\sec{SUMMARY AND DISCUSSION}

In this paper we have examined peculiar velocities of
clusters predicted in the $\Lambda$CDM model. We analyzed the
clusters in the Virgo simulation for the $\Lambda$CDM model with
$\Omega_0=0.3$, $h=0.7$ and $\sigma_8=0.9$. To identify clusters
in the simulation we used two methods: (1) the standard friends-of-friends
(FOF) method and (2) the method, where the clusters are defined as
the maxima of the smoothed density field (DMAX). We used a top-hat
window with smoothing radii $R_s=1.5h^{-1}$ Mpc and $R_s=1.0h^{-1}$
Mpc. The peculiar velocity of DMAX clusters was defined to be the mean
peculiar velocity within a sphere of radius $R_s$.

We studied the rms peculiar velocity of clusters for different
cluster masses. We found that the relation between the rms
peculiar velocity of a cluster and the mass of the cluster depends
on the method that is used to determine the cluster masses and
velocities. The rms peculiar velocity of FOF clusters decreases
as the mass of the clusters increases. The rms peculiar velocity
of DMAX clusters is almost independent of the cluster mass and
is well approximated by the linear rms velocity of peaks of the density
field smoothed at the radius $R=R_s$. We also studied the rms peculiar
velocities of peaks, $\sigma_p$, for the linear smoothing radius $R=R_L(M)$.
In this approximation, the rms velocities are significantly lower than
the rms velocities of clusters found in the simulation.

We investigated the distribution functions of the cluster
peculiar velocities for different cluster masses.
The peculiar velocity distributions found for the low-mass FOF clusters
have stronger tails than predicted by a Gaussian distribution. But the
velocity distribution of massive FOF clusters is
similar to the Gaussian distribution. The
velocity  distribution of DMAX clusters in different mass intervals is
similar. The one-dimensional velocity distribution of $R_s=1.5h^{-1}$ Mpc
clusters is well approximated by a Gaussian.
In the velocity distribution of $R_s=1.0h^{-1}$ Mpc clusters,
we found small deviations from the Gaussian distribution.

We also investigated the evolution of cluster velocities on the scale
$R_s=1.5h^{-1}$ Mpc. We found a correlation between the peculiar
velocities of massive clusters at $z=0$ and the peculiar velocities of
peaks at the initial centres of the clusters at $z=10$. Density
maxima at the scale $R_s=1.5h^{-1}$ Mpc are not isolated objects that
move with linear speeds.

In this paper we introduced the DMAX method. This method operates with
the smoothed density and velocity fields on a given scale $R_s$. We
expect that the properties of DMAX clusters (e.g. their mass distribution,
correlation function) can be determined analytically from the
properties of initial density and velocity fields, at least, for the
quasi-linear scales. The density and velocity distribution function
for the maxima in the Gaussian density field was derived by
Bardeen et al. (1986). Further study is needed to consider different
properties of the DMAX clusters.

We found that the rms velocities of DMAX clusters on a given scale
$R_s$ are well described by the linear theory at $R=R_s$
(equation 4). Suhhonenko \& Gramann (2003) studied the rms peculiar
velocities of clusters in the $\tau$CDM model. They found that also in this
model, the rms peculiar velocities of DMAX clusters are close to the
linear theory expectations. We can use the linear
approximation to estimate rms peculiar velocities of clusters in
different cosmological models.

\begin{figure}
\centering \leavevmode \psfig{file=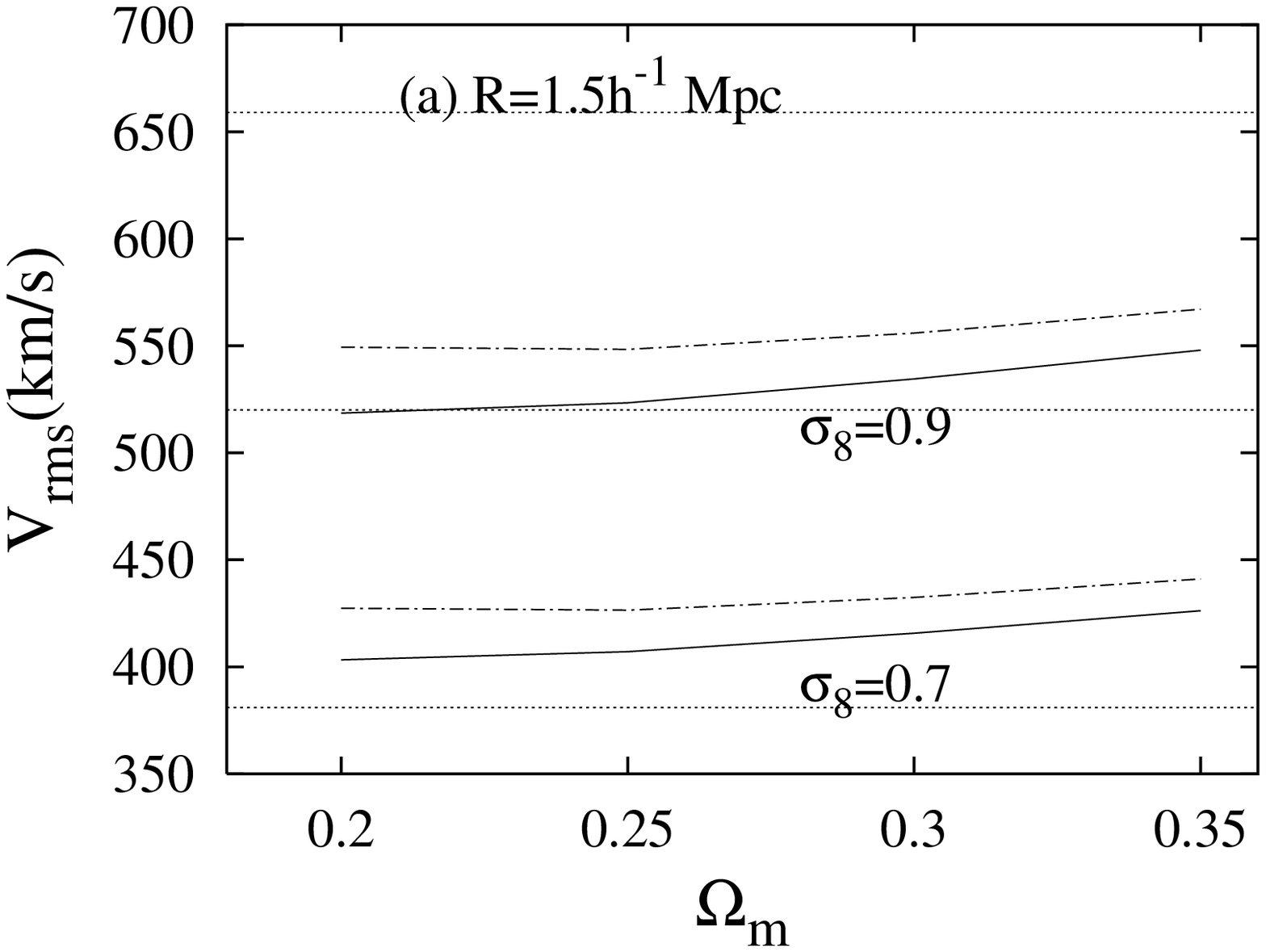,width=8cm}
\psfig{file=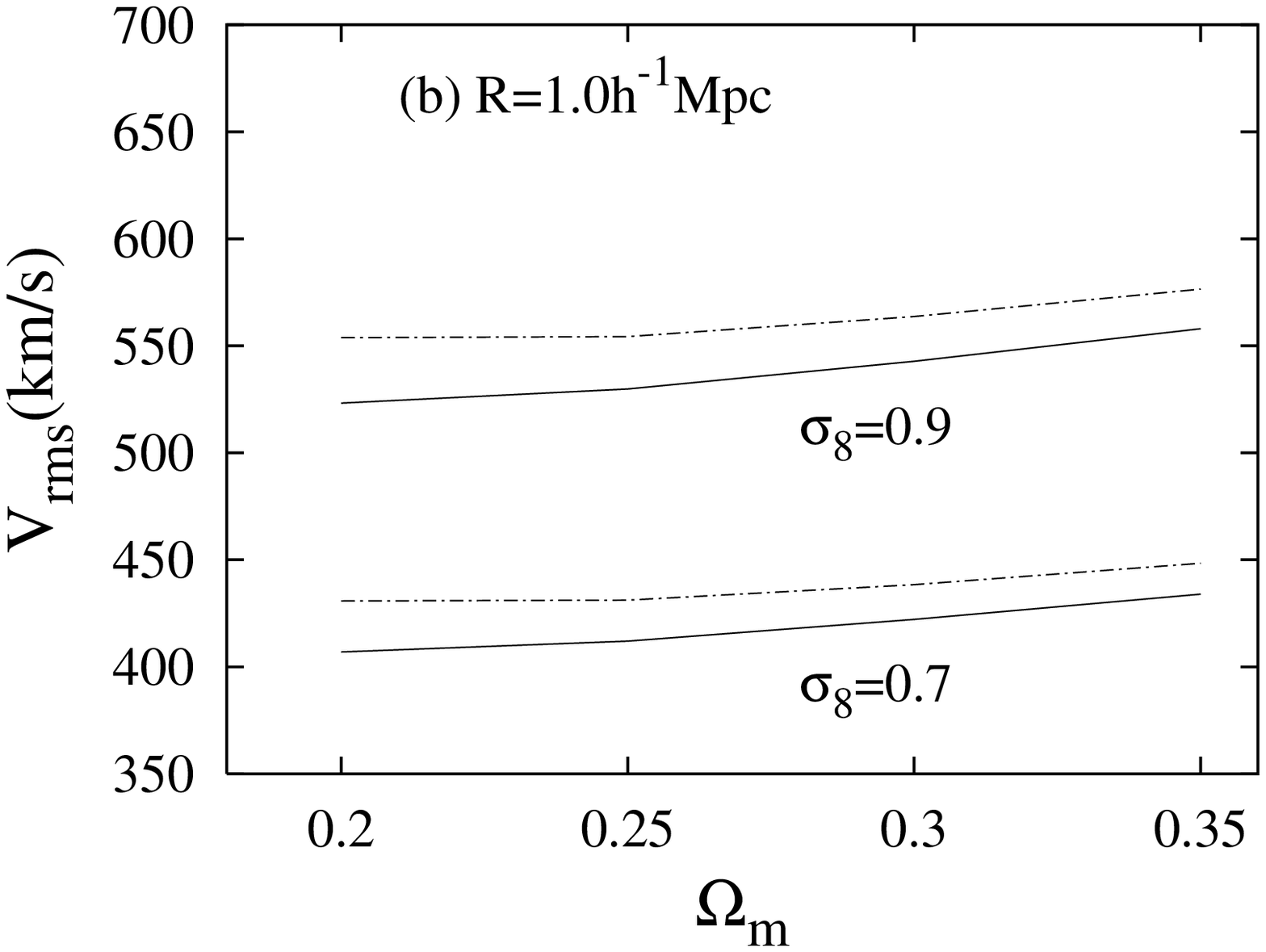,width=8cm}

\caption{The rms peculiar velocities of clusters predicted in
different flat $\Lambda$CDM models. (a) The peculiar velocities
are smoothed on the scale $R=1.5h^{-1}$ Mpc. (b) The smoothing
radius $R=1.0h^{-1}$ Mpc. The solid lines show the results for
$h=0.7$ and the dot-dashed lines for $h=0.65$. The upper curves
describe the velocities for $\sigma_8=0.9$ and lower curves for
$\sigma_8=0.7$. Dotted lines in panel (a) describe the rms
peculiar velocity obtained by Giovanelli et al. (1998) for 24
observed clusters. They found that the rms cluster peculiar
velocity is $v_{rms}=520 \pm 139 \kms$.}

\end{figure}

Fig.~6 shows the rms peculiar velocities of clusters estimated for
different $\Lambda$CDM models using the approximation (4).
Fig~6a shows the results for the radius $R=1.5h^{-1}$ Mpc and
Fig.~6b for the radius $R=1.0h^{-1}$ Mpc, respectively. We
estimated the rms peculiar velocity of clusters in flat
$\Lambda$CDM models, where the density parameter
$\Omega_m=0.2-0.35$, the normalized Hubble constant $h=0.65-0.7$
and $\sigma_8= 0.7-0.9$. The power spectrum of density
fluctuations was chosen as $P(k)=AkT^2(k)$, where the transfer
function $T(k)$ was calculated using the fast Boltzmann code
CMBFAST developed by Seljak \& Zaldarriaga (1996). The
normalization constant, $A$, was chosen by fixing the value of
$\sigma_8$. On large scales, the power spectrum
calculated by using the code CMBFAST is higher than the power
spectrum given by equation (5) (for a fixed $\sigma_8$; see, e.g.,
Fig.~2 by Gramann \& Suhhonenko 2002). To determine the
dimensionless growth rate $f$, we used the equations given by
Peebles (1984).

Consider the peculiar velocities of clusters for the smoothing radius
$R=1.5h^{-1}$ Mpc. If $h=0.7$ and $\sigma_8=0.7$, the rms peculiar
velocity of clusters is $403 \kms$ for $\Omega_m=0.2$ and $426 \kms$ for
$\Omega_m=0.35$. For a higher density parameter $\Omega_m$, the function
$f$ is larger, but the amplitude of the large-scale density
fluctuations is smaller. If $h=0.7$ and $\sigma_8=0.9$, the rms
peculiar velocity of clusters is $518 \kms$ and $548 \kms$ for
$\Omega_m=0.2$ and $\Omega_m=0.35$, respectively.

For comparison, we show in Fig.6a the rms peculiar velocity
obtained by Giovanelli et al. (1998) for 24 observed clusters
in the distance range $\sim 10$ and $90h^{-1}$ Mpc. This
comparison is very preliminary. The peculiar velocities for the
observed galaxy clusters and for the simulated DMAX clusters are
determined in different ways. Giovanelli et al. (1997) analysed
the I-band Tully-Fisher (T-F) measurements for $782$ spiral
galaxies in the fields of $24$ clusters (the SCI sample). Most of
the galaxies used in the T-F analysis were within the Abell
radius ($R_A=1.5h^{-1}$ Mpc) around the cluster centre. The
peculiar velocities for the clusters in this sample were studied
by Giovanelli et al. (1998). Individual cluster T-F relations
were referred to the average template relation to compute cluster
peculiar velocities. Giovanelli et al. (1998) found that the rms
one-dimensional cluster peculiar velocity in the SCI sample is
$300 \pm 80 \kms$, which corresponds to the three-dimensional rms
velocity $v_{rms}=520 \pm 139 \kms$. This number is in good
agreement with previous estimates by Bahcall \& Oh (1996) and by
Watkins (1997), based on the SCI sample. This estimate is also in
agreement with that determined for the SCII sample (Dale et al.
1999b). The SCII sample is based on T-F measurements for $522$
late-type galaxies in the fields of $52$ Abell clusters in the
distance range $\sim 50$ to $200h^{-1}$ Mpc. The distribution of
these galaxies in the $35$ Abell clusters was presented by Dale et
al. (1999a). Dale et al. (1999b) studied the cluster peculiar
velocities in the SCII sample and found that the rms one-dimensional
cluster peculiar velocity in this sample is $341 \pm 93 \kms$. This
corresponds to the three-dimensional rms velocity $591 \pm 161 \kms$.

Yoshikawa, Jing \& B\"orner (2003) analyzed the velocity
dispersion of galaxies, $\sigma_{gal}$, and of dark matter,
$\sigma_{dm}$, inside clusters using hydrodynamical
simulations. They found that the velocity dispersions of galaxies
in less massive ($\sim 10^{13} M_{\odot}$) clusters are
systematically lower than those of dark matter particles inside
the clusters. We do not exactly know how the mean peculiar velocity of
galaxies is related to the mean velocity of matter in the cluster.
It is usually assumed that these velocities are similar. Further
work is needed to study the peculiar velocities of galaxy
clusters in the hydrodynamical simulations.

\sec*{ACKNOWLEDGEMENTS}

We thank J. Einasto, M. Einasto, P. Heinam\"aki, G. H\"utsi,
E. Saar and S. White for useful discussions. This work has been
supported by the ESF grant 5347. The N-body simulations used in
this paper are available at {\it
http://www.mpa-garching.mpg.de/Virgo/virgoproject.html.} These
simulations were carried out at the Computer Center of the
Max-Planck Society in Garching and at the EPCC in Edinburgh, as
part of the Virgo Consortium project. The FOF programs used in
this paper are available at {\it
http://www-hpcc.astro.washington.edu.} These programs were
developed in the University of Washington.

\vfill
\end{document}